\documentclass{CFD2017}
\usepackage{CFD2017}
\pdfoutput=1
\usepackage[squaren]{SIunits}
\usepackage{subfig}

\newcommand*{\mycdot}{\kern-.2em\cdot\kern-.2em}

\newcommand{\rvec}{\textbf{r}}

\newcommand{\vvec}{\textbf{v}}
\newcommand{\uvec}{\textbf{u}}
\newcommand{\fvec}{\textbf{f}}
\newcommand{\Fvec}{\textbf{F}}

  \newcommand{\st}[1]{_{\text{#1}}} 
     \newcommand{\spt}[1]{^{\text{#1}}}

\title{Extremely fast simulations of heat transfer in fluidized beds }
\paperID{CFD 2017-024}
\author{Thomas}{Lichtenegger} 
\presenting  
\sameaddressas{1,2}
\email{thomas.lichtenegger@jku.at}
\author{Stefan}{Pirker}

\address{Department of Particulate Flow Modelling, Johannes Kepler University, 4040 Linz, AUSTRIA\newline 
$^2$Linz Institute of Technology (LIT), Johannes Kepler University, 4040 Linz, AUSTRIA}
\email{stefan.pirker@jku.at}

\begin{document}
\maketitle  
\headers   

\abstract{
  Besides their huge technological importance, fluidized beds have attracted a large amount of research because they are perfect playgrounds to investigate highly dynamic particulate flows. Their over-all behavior is determined by short-lasting particle collisions and the interaction between solid and gas phase. Modern simulation techniques that combine computational fluid dynamics (CFD) and discrete element methods (DEM) are capable of describing their evolution and provide detailed information on what is happening on the particle scale. However, these approaches are limited by small time steps and large numerical costs, which inhibits the investigation of slower long-term processes like heat transfer or chemical conversion.

In a recent study \cite{Lichtenegger2016}, we have introduced recurrence CFD (rCFD) as a way to decouple fast from slow degrees of freedom in systems with recurring patterns: A conventional simulation is carried out to capture such coherent structures. Their re-appearance is characterized with recurrence plots that allow us to extrapolate their evolution far beyond the simulated time. On top of these predicted flow fields, any passive or weakly coupled process can then be investigated at fractions of the original computational costs.

Here, we present the application of rCFD to heat transfer in a lab-scale fluidized bed. Initially hot particles are fluidized with cool air and their temperature evolution is recorded. In comparison to conventional CFD-DEM, we observe speed-up factors of about two orders of magnitude at very good accuracy with regard to recent measurements.
}
\normalfont\normalsize

\printnomenclature[0.7cm]  
\vskip .1em

\section{Introduction}
Modeling and simulation of fluidized beds is extremely demanding due to the various scales present. In the spatial domain, submillimeter particles are to be contrasted with the geometric dimensions of industrial-size plants. In the temporal domain, collisions set the lower limit of time scales at small fractions of a second while heat transfer or chemical conversion happens much more slowly.

Substantial progress has been made in up-scaling well-established mesoscopic methods like CFD-DEM \cite{Cundall1979,Tsuji1993} or the two-fluid model (TFM) \cite{Anderson1967,Gidaspow1994} to macroscopic sizes. 
To allow for coarser meshes so that larger systems can be described, sub-grid heterogeneities need to be modeled appropriately \cite{Heynderickx2004,Igci2008,Igci2011b,Igci2011,Milioli2013,Ozel2013,Parmentier2012,Radl2014,Schneiderbauer2013,Schneiderbauer2014,Wang2008,Wang2010,Yang2003,Zhang2002}.

Although industrial-size reactors may be simulated with these methods, they are still bound to short-term investigations. 
Here, we show a systematic way to circumvent this problem and apply it to heat transfer in a lab-scale fluidized bed. In a recent paper, we introduced the idea of rCFD \cite{Lichtenegger2016} to time-extrapolate the behavior of systems dominated by reappearing structures. We use fields from short-term simulations with conventional CFD-DEM or TFM to create long sequences of flow patterns with the aid of recurrence plots \cite{Eckmann1987}.

On these sequences, we simulate heat transfer between gas and solid particles as measured by Patil \textit{et al}.~\cite{Patil2015} and obtain speed-up factors of about two orders of magnitude.

\section{Model Description}
In the following, we briefly summarize the CFD-DEM method and refer the interested reader to extensive reviews \cite{Deen2007,Zhou2010}. Afterwards, the rCFD strategy is explained.
\subsection{Fluid equations}
If a secondary, particulate phase is present, the Navier Stokes equations for a fluid with density $\rho\st{f}$ and velocity field $\uvec\st{f}$ become \cite{Anderson1967}
\begin{align}
 &\frac{\partial }{\partial t}\alpha\st{f}\rho\st{f} + \nabla \cdot \alpha\st{f}\rho\st{f} \uvec\st{f} = 0 \label{eq:gaseom1}\\
 &\frac{\partial }{\partial t}\alpha\st{f}\rho\st{f}\uvec\st{f} + \nabla \cdot \alpha\st{f}\rho\st{f} \uvec\st{f}\uvec\st{f} =
  -\alpha\st{f}\nabla p\st{f}+
  \alpha\st{f}\nabla\cdot\boldsymbol\tau\st{f} + \fvec\st{drag} + \fvec\st{ext}.\label{eq:gaseom2}
\end{align}
Various correlations for particle-fluid drag $\fvec\st{drag}$ can be found in literature, we use one obtained from lattice-Boltzmann simulations~\cite{Beetstra2007}.

To picture heat transfer, we assume incompressible conditions and neglect contributions due to pressure variations to the enthalpy transport equation from which we derive
\begin{align}
 C\spt{(p)}\st{f}\frac{\partial }{\partial t}\alpha\st{f}\rho\st{f}T\st{f} + C\spt{(p)}\st{f}\nabla\cdot \alpha\st{f}\rho\st{f}\uvec\st{f}T\st{f}
  = \nabla\cdot \alpha\st{f}k\st{f}\spt{(eff)} \nabla T\st{f} + \dot{Q}\st{p-f}.\label{eq:gaseom3}
\end{align}
Both heat conduction $\nabla\cdot \alpha\st{f}k\st{f}\spt{(eff)} \nabla T\st{f}$ and particle-fluid heat transfer $\dot{Q}\st{p-f}$ have to be modeled with empirical correlations \cite{Syamlal1985,Gunn1978}. Note that for incompressible flows, Eq.~\eqref{eq:gaseom3} is decoupled from Eqs.~\eqref{eq:gaseom1} and \eqref{eq:gaseom2} if density and viscosity are assumed to be temperature-independent.

\subsection{Particle equations}
Solid particles are often modeled as perfect spheres which interact with each other via contact forces $\Fvec_i\spt{(p-p)}$, with a surrounding fluid $\Fvec_i\spt{(p-f)}$ and possibly other sources $\Fvec_i\spt{(ext)}$.
For each particle $i$ with mass $m_i$, Newton's second law
\begin{align}
 m_i\dot{\vvec}_i = \Fvec_i\spt{(p-p)} + \Fvec_i\spt{(p-f)} + \Fvec_i\spt{(ext)} \label{eq:parteom1}
\end{align}
determines its trajectory.
The particle-particle force 
\begin{equation}
 \Fvec_i\spt{(p-p)} = \sum_{j\neq i} \big(\Fvec\spt{(n)}_{i,j} + \Fvec\spt{(t)}_{i,j}\big)
\end{equation}
on particle $i$ due to all surrounding particles $j$ consists of normal and tangential components which depend on the relative positions and velocities of $i$ and $j$ and are often described via spring-dashpot models \cite{Cundall1979}. The interaction of particles with a fluid $\Fvec_i\spt{(p-f)}$ is related to the drag force and the pressure gradient in Eq.~\eqref{eq:gaseom2}.

Each particle is assumed to have a homogeneously distributed temperature $T_i$. We neglect particle-particle heat exchange because of the extremely short collision times so that only heat transfer from/to the fluid can change a particle's temperature via
\begin{equation}
 m_i C\st{p}\spt{(p)}\dot{T}_i = -\pi k\st{f} D_i \text{Nu}\st{p} (T_i - T\st{f}(\rvec_i))\label{eq:parteom3}
\end{equation}
in terms of the Nusselt number $\text{Nu}\st{p}$ \cite{Gunn1978}.

\subsection{Recurrence CFD}
In a first step, we define norms for recurrence and dissimilarity of states, respectively, e.g.\
\begin{align}
& R(t,t')  \equiv 1-\frac{1}{{\cal N}}\int d^3r \big(\alpha\st{f}(\rvec,t) - \alpha\st{f}(\rvec,t')\big)^2\label{eq:recmatdef}\\
&  D(t,t')\equiv 1 - R(t,t') \label{eq:recmatdef2},
\end{align}
to assess the similarity of flow fields obtained from the solution of Eqs.~\eqref{eq:gaseom1} -- \eqref{eq:parteom1} at two times $t,t'\leq t\st{max}$.
\begin{equation}
 {\cal N} \equiv \text{max}_{t,t'} \int d^3r \big(\alpha\st{f}(\rvec,t) - \alpha\st{f}(\rvec,t')\big)^2
\end{equation}
ensures normalization in Eq.~\eqref{eq:recmatdef}. We stress that it is an assumption that similarity according to Eq.~\eqref{eq:recmatdef} carries over to other fields like the particle velocity $\uvec\st{p}$. As a matter of fact, it has to be checked \textit{a posteriori} if they are sufficiently correlated to justify it. 

Compared to binary recurrence statistics \cite{Eckmann1987}, definition \eqref{eq:recmatdef} allows for continuous degrees of similarity, which is referred to as unthresholded recurrence statistics \cite{Marwan2007}.

\begin{figure}[htbp]
\centering
\subfloat[\label{fig:v1}]{%
\includegraphics[width=0.3\textwidth]{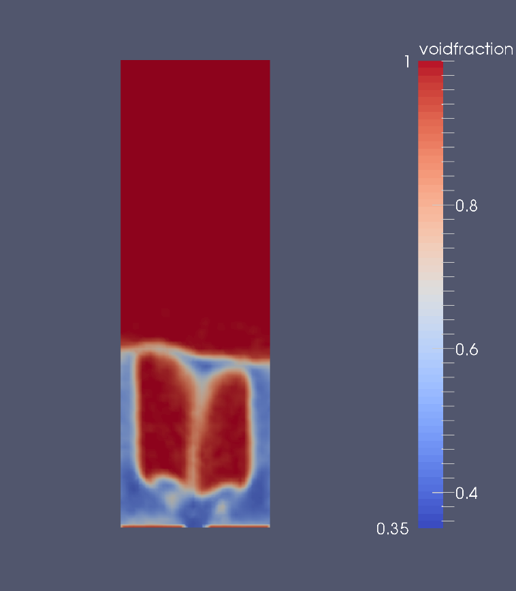}
}\\
\subfloat[\label{fig:v2}]{%
  \includegraphics[width=0.3\textwidth]{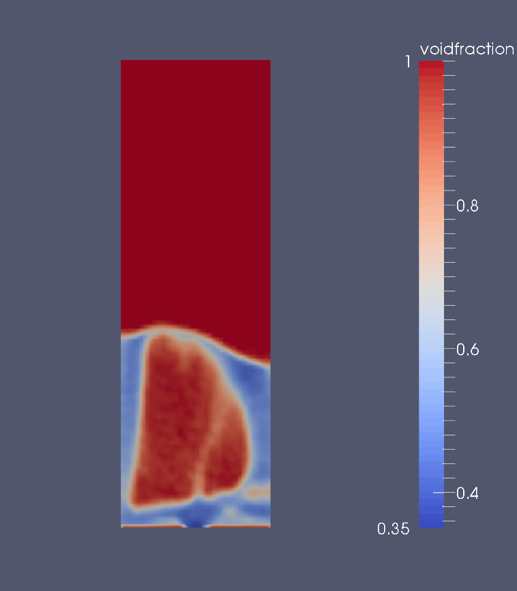}
}
\caption{Example of two equivalent states in the sense of dissimilarity norm Fig.~\ref{fig:recplot}. Although the volume fraction fields are not identical, they are close enough so that substituting one with the other does not lead to any significant inaccuracies.}
	\label{fig:voidfraction}
\end{figure}

Given a large enough recurrence statistics $R(t,t')$, it is possible to extrapolate the underlying system's behavior.
Starting at some given begin time $t_i\spt{(b)}$, one randomly picks an interval of length $\Delta t_i$. In this study, we used uniformly distributed $\Delta t_i$ in the range of $t\st{max}/20$ and $t\st{max}/5$. 
Within $\Delta t_i$, the corresponding fields are taken as first elements of the desired sequence. Then, the most similar state to that at the end of the interval $t_i\spt{(e)} = t_i\spt{(b)} + \Delta t_i$ is identified with the aid of $R(t,t')$ and reconstruction is continued from there on. Again, an interval length is chosen and its fields are appended to the sequence. Obviously, this procedure can be repeated arbitrarily often. Figure~\ref{fig:voidfraction} shows an example of two states that could be identified with each other because of a sufficiently high degree of similarity.
\InsFig{figs/recplot}{Degree of dissimilarity $D(t,t')$ for the first $5\,\second$ of simulation of a fluidized bed corresponding to 200 snapshots. Besides the main diagonal with dissimilarity $0$, a pattern of local minima and maxima is clearly visible, corresponding to more or less similar states. Their alignment approximately parallel to the main diagonal indicates pseudo-periodic behavior.}{recplot}

Finally, any passive processes can now be simulated on these fields in an extremely efficient way.
The CFD-side problem of Eqs.~\eqref{eq:gaseom1}, \eqref{eq:gaseom2} and \eqref{eq:gaseom3} is replaced with the temperature equation \eqref{eq:gaseom3} alone. Other information like volume fraction, density and velocity are obtained from the recurrence process, only the temperature distribution is really calculated.
The motion of the particles is simplified with tracers that follow the solid phase's velocity field $\uvec\st{p}\spt{(rec)}$ and undergo random fluctuations $d\textbf{w}_i$, viz.
\begin{equation}
d\rvec_i=\uvec\st{p}\spt{(rec)}(\rvec_i,t)dt + \sqrt{2D\spt{(rec)}(\rvec_i,t)}d\textbf{w}_i.\label{eq:recparteom1}
\end{equation}
The latter act to avoid too high tracer concentrations in comparison to actual solid particles. A phenomenological, possibly position-dependent parameter $D\spt{(rec)}$ controls the strength of the fluctuations, e.g.\ via
\begin{equation}
 D\spt{(rec)}(\rvec,t) = D_0 \frac{\text{max}\Big[\alpha\st{p}(\rvec,t) - \alpha\st{p}\spt{(rec)}(\rvec,t),0\Big]}{\alpha\st{p}(\rvec,t)}.
\end{equation}
The constant $D_0$ has to be chosen empirically such that neither excessive clustering occurs nor that particle diffusion is enhanced dramatically. In neither case, the passive process under consideration could be pictured accurately.

\section{Simulation setup}
 We set up simulations in close resemblance to recent experiments and simulations of heat transfer in a fluidized bed \cite{Patil2015} which we took as reference data. A $8 \, \centi\meter$ $\times$ $1.5 \, \centi\meter$ $\times$ $25 \, \centi\meter$ cuboid was discretized into $35 \times 110 \times 6$ equal cells and approximately {57000} $1$-mm spheres with material values of glass and initial temperature $T\st{p}^{(0)} = 90\,\celsius$ were inserted. Air with ambient temperature $T\st{f}^{(0)} = 20\,\celsius$ fluidized them with a superficial inlet velocity of $u\st{inlet} = 1.2\, \meter/\second$ and slowly cooled them.
 
Since unresolved CFD-DEM requires larger cells than particles, the above grid could not be used to resolve the flux of heat through the domain walls happening over a very thin layer. Instead, its thickness was introduced as modeling parameter within reasonable bounds to control heat loss through the walls \cite{Patil2015}. Since the focus of this work was rather the comparison of full CFD-DEM with rCFD than with experiments, we chose its value (in roughly the same range, but somewhat larger) such that our CFD-DEM results were in agreement with the reference simulation. We opted for this procedure because in contrast to the reference simulation which was carried out for an ideal, compressible gas, we approximated the system as incompressible to enable a recurrence-based treatment. Future work will address this shortcoming \cite{Lichtenegger2017} and allow for a meaningful comparison with measurements.

For the full CFD-DEM simulations, Eqs.~\eqref{eq:gaseom1} -- \eqref{eq:gaseom3} were solved under incompressible conditions employing the PISO algorithm \cite{Issa1986}. On the DEM side, we used velocity-Verlet integration \cite{Verlet1967} for Eq.~\eqref{eq:parteom1}. The full CFD-DEM simulations were carried out for $10\,\second$ process time, the first $5\,\second$ were then used as basis for rCFD. We stress that this duration was chosen quite generously and smaller recurrence statistics would probably work equally well.

In this study, each particle was represented by exactly one tracer. Of course, one could also employ a parcel approach to combine clusters of particles into fewer tracers or conversely try to resolve coarse-grained CFD-DEM data with higher numbers of tracers.

Special care had to be taken with the choice of the fluctuation control parameter $D_0$. We found empirically that with $d\textbf{w}$ distributed uniformly on the unit sphere, 
$D_0 = 10^{-4}\, \meter^2/\second$ was large enough to avoid too high particle concentrations and sufficiently small not to increase particle diffusion and consequently heat transfer and the mean cooling rate.

\section{Results}
According to the rCFD strategy outlined above, the first quantity to look at is the dissimilarity norm (or equally useful the recurrence norm) displayed in Fig.~\ref{fig:recplot}. 200 snapshots corresponding to $5\,\second$ real time are compared with each other, leading to a structure of local minima and maxima approximately parallel to the main diagonal. These minima are caused by very similar states that evolve in an analogous fashion, which demonstrates local pseudo-periodicity.

The degree of similarity is indicated in Fig.~\ref{fig:voidfraction}. Two volume fraction fields which give rise to a local minimum in the dissimilarity norm are shown. While small quantitative differences can be found, their qualitative features clearly agree: a large bubble was formed in the center of the bed with a few particles moving downwards in the middle of the bubble.
\InsFig{figs/temp}{Particle mean temperature over time. Per construction, the full CFD simulations closely resembled the reference data \cite{Patil2015}. As a consequence, the cooling rate was somewhat too low in comparison with measurements. Most importantly, recurrence CFD led to almost identical results as the full model.}{temp}

If such semiquantitative agreement was sufficient to obtain a meaningful extrapolation of the fluidized bed's evolution is finally answered by looking at the particle mean temperatures from full and recurrence-based simulations in Fig.~\ref{fig:temp}. As discussed above, the thickness of the boundary layer over which temperature drops to its wall value allowed for some ``tuning'' of the cooling rate. With a value to match the reference data \cite{Patil2015}, the agreement of the present CFD-DEM results with them was of course to be expected. More importantly, however, the curves from rCFD and CFD-DEM agree very well. After $10\,\second$ their deviation is much less than $1\,\celsius$ at a speed-up of approximately two orders of magnitude.

Furthermore, Fig.~\ref{fig:temp} demonstrates that particle cooling happened over much longer times than the bed's fast dynamics. Such a clear separation of scales is clearly vital for the rCFD procedure because it minimizes the impact of switching between most similar but nevertheless different flow states on the passive process under consideration.

\section{Conclusion and outlook}
In this paper, we have demonstrated how to decouple the fast dynamics of a fluidized bed from the much slower heat transfer between solid particles and surrounding gas to accelerate the calculations. To facilitate the procedure, we assumed incompressible conditions in contrast to the treatment in the reference simulation \cite{Patil2015} and used a tuning parameter for wall heat loss to match our full CFD-DEM simulations to them. Upon these results, we built the recurrence statistics used for rCFD.

Future work \cite{Lichtenegger2017} will deal with rCFD for the compressible case, where we expect a superposed transient behavior due to a slow decrease of the mean gas temperature in the center of the bed. Besides more realistic results from the CFD-DEM calculation for a meaningful comparison with measurements, this will allow to study the bed over much longer times. Furthermore, a detailed performance analysis of the method's implementation might reveal optimization potential for even faster simulations.

\section{Acknowledgement}
This work was partly funded by the Linz Institute of Technology (LIT), Johannes Kepler University, Linz. We furthermore acknowledge support from K1-MET GmbH metallurgical competence center.

\bibliographystyle{CFD2017}
\bibliography{References}

\end{document}